\documentclass[aps,prb,groupedaddress,twocolumn,10pt]
{revtex4-1}

\usepackage{graphicx}

\begin{document}

\newcommand{\be}{\begin{equation}}
\newcommand{\ee}{\end{equation}}
\newcommand{\bea}{\begin{eqnarray}}
\newcommand{\eea}{\end{eqnarray}}

\title{Searching for non-Fermi liquids under the holographic light}

\author{D. V. Khveshchenko}

\affiliation{Department of Physics and Astronomy, University of North Carolina, Chapel Hill, NC 27599}

\begin{abstract}
By expanding the set of background geometries beyond the commonly studied ones 
we identify those dual gravity models 
that may provide holographic descriptions for some prototypical non-Fermi liquid 
states of strongly correlated condensed matter systems.
Specifically, we discuss prospective gravity duals of such iconic examples as the 
non-relativistic fermions coupled to gauge fields and Dirac fermions with the Coulomb interactions.
\end{abstract}

\maketitle

\nopagebreak
Quantum theory of strongly correlated fermions has long been in 
a rather desperate need for non-perturbative techniques, the use of 
which could allow one to proceed beyond the customary (and often 
uncontrollable in the regime of interest) approximations 
when analyzing generic (non-integrable) systems.

Despite all the effort, though, the overall progress 
has been quite limited. However, it's been argued that a recent 
proliferation of the ideas based on the conjecture of holographic 
duality \cite{AdS} may offer a possible way out of the stalemate,
thus allowing the field to move towards a systematic classification of various 
'strange' metallic (compressible) states that are commonly (and often indiscriminately) 
referred to as non-Fermi-liquids (NFLs).

The holographic correspondence postulates a connection 
between certain (non-Abelian, multi($N$)-component, and supersymmetric) 
field theory models and their gravity duals living in one extra dimension,
so that for $N\gg 1$ the strong-coupling regime of the former can be 
mapped on the weak-coupling one of the latter (and vice versa).
The support for this general idea is provided by a host of circumstantial
evidence gathered from the 'bona fide' theories 
of strings and hot QCD quark-gluon plasmas.  

Nevertheless, despite a gradually building confidence in the validity of 
the original holographic conjecture, the status 
of its recently proposed condensed matter 
applications remains, by and large, unknown.

Pursuing the phenomenological 'bottom up' approach, the initial studies  
produced a number of rather baffling results, 
which include multiple Fermi surfaces (merging into one critical 
'Fermi ball' - or, rather, 'flat band'- in the extreme $N\to\infty$ limit), oscillatory frequency dependence and dispersionless 
poles of the fermion propagator, etc. \cite{AdS2}. Some of those early findings have already been interpreted as spurious artefacts of taking the limit $N\to\infty$, though. 
For one, no oscillatory dependence would arise in a more 
systematic 'top down' approach \cite{topdown} (besides, the same effect can also achieved by including a non-minimal (Pauli) fermion-gauge field coupling \cite{phillips}).

Putting aside the central question about a general applicability of the original holographic 
conjecture to (typically, neither non-Abelian/multicomponent, nor supersymmetric) 
condensed matter systems, one common limitation of the early studies 
was that their background metrics would typically be chosen from 
a handful of the well-known solutions to the classical Einstein-Maxwell equations.

Obviously, such an 'under the light' search for holographic NFLs lacks any physical input specific to a given strongly correlated system
and, therefore, its chances of finding a gravity dual for that particular 
system appear to be rather hard to assess. 

In essence, the early studies have so far found just one type of the NFL behavior, dubbed 'semi-local criticality', whereby the fermion 
propagator features a non-analytical (and, in general, oscillatory) 
frequency dependence, but only a non-singular momentum one
\be
G(\omega, q)={1\over A(q)+B(q)\omega^{\nu_q}}
\ee
Here the function $A(q)$ has simple zeros at each of the 
(potentially, multiple) 'Fermi momenta' $q_F^i$, 
whereas $B(q)$ takes finite values at such points (see Refs.\cite{AdS2} for details). 

This behavior suggests that at long distances 
the system effectively splits onto spatially uncorrelated  
'quantum impurities', each of which exhibits a characteristic 
$d=0$ quantum-critical scaling.

Although it was pointed out that the propagator (1) bears a certain resemblance to that expected in the context of some heavy fermion materials (see, e.g., \cite{heavyfermion} and references therein), in the absence of any solid agreement with experiment that would suggest otherwise, this reminiscence might well turn out to be merely superficial.

In view of such uncertainty, the task of putting the holographic correspondence on a firm ground and ascertaining the status of its predictions more definitively
would be best achieved if this technique were applied 
to those situations where a prior insight has already been gained by some other means. To that end, it would be very helpful first to find 
the gravity duals of the already established (or, at least, suspected) NFL states.

In what follows, we demonstrate that accomplishing this task  
requires one to extend the class of metrics well beyond
the commonly studied classical examples. 

{\it Prospective gravity duals of condensed matter systems}

The early applications of the holographic conjecture yielding Eq.(1)
utilized the standard Reissner-Nordstrom (RN) 'black brane' solution 
which minimizes the Einstein-Maxwell action
\be
S_g=\int{1\over 2\kappa^2}({\cal R}+{d(d+1)\over L^2})-{1\over 2e^2}F_{\mu\nu}^2
\ee
Hereafter $\int=\int dt dz d^d{\vec x}{\sqrt {-det g_{\mu\nu}}}$ 
stands for a covariant $d+2$-dimensional 
volume integral, $\cal R$ is the scalar curvature, the second term represents 
a (negative) cosmological constant of the asymptotically 
anti-de-Sitter space $AdS_{d+2}$ of curvature radius $L$, $\kappa^2$ is the Newtonian coupling, and $e$ is the charge of the $U(1)$ gauge field.  

The corresponding metric (throughout this paper, the speed of light $c=1$)
\be
ds^2=-f(z)dt^2+g(z){dz^2}+h(z)d{\vec x}^2
\ee
has non-zero components $f(z)(z/L)^2={(L/z)^2/g(z)}=1-(1+\mu^2)
({z/z_h})^{d+1}+\mu^2({z/z_h})^{2d}, h(z) = ({L/z})^2$.
Here $\mu$ is the (dimensionless) chemical potential of the fermion matter 
which is assumed not to react back on the metric, 
and $z_h$ is the inverse radius of the horizon determined from $f(z_h)=0$. 

The latter can remain finite even when the Hawking temperature  
$T=(d+1-(d-1)\mu^2)/4\pi z_h$ vanishes, 
thereby giving rise to yet another artefact of the $N\to\infty$ limit, 
a seemingly non-vanishing entropy $S(T\to 0)\neq 0$. 
In the near-boundary $z\to 0$ (ultraviolet or UV) regime, Eq.(3) 
recovers the standard $AdS_{d+2}$ form, $f(z)=g(z)=h(z)=(L/z)^2$.

In fact, the 'locally critical' behavior (1) sets in at the extremal $T\to 0$ limit
\cite{AdS2}, where the near-horizon geometry approaches $AdS_2\times R^d$.
Moreover, a similar (albeit more physically sound, entropy-wise: $S(T)\sim T^{d/\eta}$) 
behavior was found for a variety of geometries
which reduce to the one-parameter 'Lifshitz' metric
 $f(z)=(L/z)^{2\eta},~~~g(z)=h(z)=(L/z)^2$ in the $z\gg 1$ 
(infrared or IR) regime \cite{lifshitz}. It was also shown to result from an approximate (Thomas-Fermi) account of the fermions' back-reaction in the framework of the standard gravity (2), thus leading to the 'electron star' scenario \cite{electronstar}.

In fact, the latter metric naturally emerges alongside a whole class of more general solutions in the so-called dilaton gravity whose Lagrangian includes an additional bulk scalar field 
\cite{dilaton} 
\be
S_{dg}=\int{1\over 2\kappa^2}({\cal R}-{(\partial\phi)^2\over 2}+U(\phi))
-{Z(\phi)\over 2e^2}F_{\mu\nu}^2
\ee
In the minimal version, both, the dilaton potential $U(\phi)=d(d+1)e^{\delta\phi}/L^2$ and the effective gauge coupling $Z(\phi)=e^{\gamma\phi}$, are given by simple exponential functions    
with the coefficients $\delta$ and $\gamma$. 

In what follows, we focus on the $T=0$ case 
and consider a still broader class of static and spherically symmetric metrics 
\begin{equation}
f(z)=(L/z)^{2\eta},~~~g(z)=(L/z)^{2\alpha},~~~h(z)=(L/z)^{2\beta},
\end{equation}
whereas at finite $T$ one also has the freedom of altering the additional polynomial 'emblackening factor', similar to that in the RN solution (3).

For any $\beta\neq 0$ Eq.(5) can be reduced to a two-parameter family of
metrics known as the 'hyperscaling violating' backgrounds 
\cite{hyperscaling}.
The latter are characterized by the dynamical 
exponent $\zeta$ and hyperscaling violation parameter $\theta$ 
\be 
\zeta={\eta+1-\alpha\over 1-\alpha+\beta},~~~~\theta=d{1-\alpha\over 1-\alpha+\beta}
\ee
which manifest themselves through the scaling properties of the excitation spectrum in the boundary theory: $\omega\to\lambda^\zeta\omega$ for $q\to\lambda q$ 
and that of the interval (3), $ds\to \lambda^{\theta/D} ds$.  

In Refs.\cite{hyperscaling}, a number of 'top down' 
string scenaria was presented, 
by which the 'hyperscaling violating' geometries may arise. 
In that regard, the null energy criteria that must be obeyed 
by any physically sensible gravity dual with the metric (5) impose the inequalities
\be
\beta(\eta-\beta+\alpha-1)\geq 0,~~~(\eta-\beta)(1-\alpha+\eta+d\beta)\geq 0
\ee
As in Refs.\cite{hyperscaling}, the generalized metrics (5) do not cover the $z\to 0$ region
and, therefore, require a proper UV completion. Thus, they should be viewed as 
gravity duals of some effective IR field theories residing at a finite $z_0$.
Correspondingly, all the holographic propagators discussed in the rest of this paper pertain to the renormalized operators from such effective theories, rather than those of 
the 'microscopic' boundary ones. The latter can be obtained 
from the former by virtue of the matching procedure, akin to that of Refs.\cite{AdS2}. 
 
{\it Semiclassical propagators}

In the holographic analyses the bulk fermions of mass $m$ 
couple to the metric and gauge fields in the minimal way
\be
S_f=\int {\bar \psi}\gamma_\mu(i\partial_\mu +{i\over 8}[\gamma_\lambda,\gamma_\nu]\omega^\mu_{\lambda\nu}+eA_\mu-m)\psi
\ee
where $\gamma_\mu$ are the $\gamma$-matrices and $\omega^\lambda_{\mu\nu}$ is the spin connection \cite{AdS}. 

In the absence of explicit analytical solutions for the bulk fermion wavefunctions in generic gravitational backgrounds, one can still resort to the semiclassical 
approach. The equation for the Fourier-transformed wavefunction
$\psi(z,\omega,q)$ features an effective single-particle potential 
\cite{WKB_k}
\be
V(z)=m^2+{q^2\over h(z)}+{\omega^2\over f(z)}
\ee
which allows for two zero-energy solutions in the tunneling region $z_0<z<z_t$:
\be
\psi_{\pm}(z,\omega,q)\sim {1\over V^{1/4}(z)}
e^{\pm\int^{z_t}_{z_0}dz{\sqrt {g(z)V(z)}}}
\ee
where the turning point $z_t$ is defined as $V(z_t)=0$. 

Using Eq.(10) one then finds the effective IR theory's Green function 
as the reflection coefficient for the 
wave incident at $z=z_0$ \cite{AdS}
\be
G_{IR}(\omega,q)={\psi_-(z,\omega,q)\over \psi_+(z,\omega,q)}|_{z\to z_0}\sim e^{-S(\omega,q)}
\ee
where
\be
S(\omega,q)=2\int^{z_t}_{z_0} dz {\sqrt {g(z)V(z)}}
\ee
Considering the metric (5) and focusing on the limit of a small 
fermion mass, one obtains the scaling behavior 
\be
S(\omega, q)\sim({q^{1-\alpha+\eta}\over \omega^{1+\beta-\alpha}})^{1\over \eta-\beta}
\ee
indicative of the underlying quasiparticle dispersion $\omega\sim q^\zeta$ governed by the dynamical exponent (6).

In the complementary limit of a large mass, the semiclassical analysis can be more conveniently employed directly in the real space \cite{WKB_x}.
In this regime, various quantum-mechanical amplitudes are dominated
by the fermion paths that closely follow the classical 
trajectories (geodesics) obtained from the (imaginary-time) action
\be
S(\tau,x)=m\int dz {\sqrt {g(z)+f(z)({d\tau/dz})^2+h(z)({d x/dz})^2}} 
\ee 
When evaluated upon such a trajectory, Eq.(14) reads 
\be
S(\tau, x)=m\int^{z_t}_{z_0} dz{\sqrt {g(z)\over r(z)}}
\ee
where $r(z)=1-\Pi^2_x/h(z_t)-\Pi^2_\tau/f(z_t)$ 
is a function of the conjugate momenta $\Pi_x=\delta S/\delta (dx/dz)$ and
$\Pi_\tau=\delta S/\delta (d\tau/dz)$
given by the integral equations
\be
x=\Pi_x
\int^{z_t}_{z_0} {dz\over h(z)}{\sqrt {g(z)\over r(z)}},~~~
\tau=\Pi_\tau
\int^{z_t}_{z_0} {dz\over f(z)}{\sqrt {g(z)\over r(z)}}
\ee
and the turning point is obtained by solving the equation $r(z_t)=0$. 
The minimal action (15) then controls the fermion propagator,
$G(\tau, x)\sim e^{-S(\tau, x)}$.

While an explicit computation of Eq.(15) can only be performed in 
some special cases, determining simpler, one-parameter, 
dependences $S(\tau)$ and $S(x)$ is possible for a broad variety of metrics.
Specifically, for the metric (5) one obtains
\be
S(x)\sim x^{1-\alpha\over 1-\alpha+\beta},~~~
S(\tau)\sim \tau^{1-\alpha\over 1-\alpha+\eta}
\ee
Notably, both Eqs.(12) and (15) elucidate the role of the radial  
variable $z$ as an energy-like renormalization scale parameter.  
However, a direct correspondence between the two can not be readily established, 
since the Fourier transformation relating $G(\omega,q)$
and $G(\tau,x)$ requires both functions, including their non-exponential prefactors, 
to be known across the entire ranges of their arguments.
It might be possible, though, to relate 
their asymptotics by virtue of the saddle-point method, wherever applicable. 

{\it Finite density fermions with singular interactions}

One important testing ground for the holographic conjecture is provided 
by the theory of finite density fermions coupled to an Abelian gauge field.
This problem has long been at the forefront of theoretical research
where it was studied with a whole variety of techniques, although the case 
still remains unclosed.
For instance, the recent results of Refs.\cite{gauge} which revisited the attempts to obtain a self-consistent re-summation to all orders in the spirit of the Eliashberg theory \cite{aim} indicate that a naive $1/N$-expansion may not be as reliable as previously thought.

Furthermore, long-ranged and retarded ('singular')
interactions that allow for a similar description are often associated with the onset of ground state instabilities, and the concomitant NFL behaviors might occur even in those systems whose microscopic Hamiltonians involve only short-ranged couplings.

Such interactions are mediated by gapless bosonic excitations 
of an emergent order parameter, and in all the diverse 
reincarnations of the problem their gauge-like propagator
conforms to the general expression 
\be
D(\omega,q)={1\over |\omega|/q^\xi+q^\rho}
\ee
Important pertinent examples include anomalous electromagnetic skin effect in metals
\cite{reizer},
compressible Quantum Hall states with screened repulsive interactions \cite{hlr},  
critical spin fluctuations in itinerant ferromagnets \cite{hertz} and density 
fluctuations in 'quantum nematics' \cite{nematics}, 
for all of which $\xi=1, \rho=2$.
In contrast, normal skin effect and antiferromagnetic fluctuations 
would be described by $\xi=0, \rho=2$,
whereas compressible Quantum Hall states with the 
unscreened Coulomb interactions correspond to $\xi=1, \rho=1$. 

The asymptotic IR behavior of the
propagator of fermions coupled to a gauge-like bosonic mode can be evaluated
by means of the eikonal-type procedure \cite{eikonal}
which reduces the former to the phase factor taken along the classical trajectory 
\be
G(\tau, x)\sim <\exp(i\int A_\mu(z=z_0)dx_\mu)>_A=e^{-S}
\ee
Here the averaging is performed over a (physical or effective) gauge field $A_\mu$
governed by the propagator $<A_\mu A_\nu>=D(\omega,q)
(\delta_{\mu\nu}-q_\mu q_\nu/q^2)$, thereby resulting in 
\be
S(\tau,x)={1\over 2}\int{d\omega d^dq\over (2\pi)^{d+1}}
D({\omega,q}){1-\cos(\omega\tau-{\bf q}{\bf x})\over (i\omega-{\bf v}{\bf q})^2}
\ee
where ${\bf v}\sim q_F{\bf {\hat x}}$ is the Fermi velocity in the direction of the vector $\bf x$.

To estimate the eikonal action (20) for a time-like interval
\be
S(\tau)={1\over 2}\int{d\omega d^dq\over (2\pi)^{d+1}}
D({\omega,q}){1-\cos\omega\tau\over (i\omega-{\bf v}{\bf q})^2}
\sim\tau^{\rho+1-d\over \xi+\rho}
\ee
we first perform the momentum and then the frequency integrations,
thereby discovering that at $\tau\to\infty$ the integral is dominated by the frequencies $\omega\sim\tau^{-1}$ and momenta $q\sim\tau^{-1/\xi+\rho}$.

Moreover, the kinematics of fermion scattering is such that 
at small scattering angles (which is the regime amenable to the eikonal approximation) one finds $\omega\ll |{\bf v}{\bf q}|\ll |{\bf v}\times{\bf q}|$ for all $2-\xi<d<1+\rho$. 
Thus, the scattering momentum appears to be primarily directed 
along the Fermi surface and perpendicular to the local Fermi velocity. 

With that observation in mind one can also compute 
the integral (20) for a space-like interval,
thus obtaining the asymptotic behavior 
\be
S(x)\sim x^{\rho+1-d\over \xi+d-1}
\ee
It is then easy to see that the Eqs.(17) and (21,22) match, provided that the following relations hold
\bea
{1-\alpha\over 1-\alpha+\beta}={\rho+1-d\over \xi+d-1}\nonumber\\
{1-\alpha\over 1-\alpha+\eta}={\rho+1-d\over \rho+\xi}
\eea
The above results pertain to the propagators of the effective field theories 
which, unlike their underlying microscopic counterparts, 
become essentially universal after having undergone
renormalization down to the IR scale $z_0$. As such, they need to be contrasted
with the holographic Green functions 
computed at a finite $z_0$, rather than those at the original boundary $z=0$. 

Thus, in order to reproduce the effects of the singular interaction 
(18) with $\xi=1,\rho=2$ and $d=2$ in the holographic setting, one needs
to choose the relevant parameters as follows: $\eta=2\beta=2(1-\alpha)$.
For comparison, the case of $\xi=1, \rho=1$ can be covered by choosing  
$\eta=\beta, \alpha=1$, whereas the case of $\xi=0, \rho=2$ requires 
$\eta=1-\alpha, \beta=0$, which values would be unattainable within 
the class of hyperscaling violating metrics. 

Notably, all of the above metrics comply with the criteria 
(7) for the existence of a consistent gravity dual, 
the second one being satisfied as a strict  
equality, $\eta=1-\alpha+\beta$ or $\zeta= 1+\theta/d=(2(1-\alpha)+\beta)/(1-\alpha+\beta)$. Also,  
the corresponding values of the dynamical exponent ($\zeta=3/2, 1$, and $2$, respectively) agree with those inferred from contrasting the quasiparticle dispersion ${\bf v}{\bf q}$ against the fermion self-energy \cite{gauge,aim,reizer,hlr,hertz,nematics} 
\be
\Sigma(\omega)=\int{d\epsilon d^dq\over (2\pi)^{d+1}}
{D(\omega,q)\over i\omega+i\epsilon-{\bf v}{\bf q}}\sim\omega^{d-1+\xi\over \xi+\rho} 
\ee
which comparison yields $\zeta=(\xi+\rho)/(d-1+\xi)$.

Obviously, the matching conditions (23) are only the necessary ones, so they may not 
always guarantee that the entire two-parameter functional dependence of the action $S(\tau,x)$
would be reproduced with this choice of parameters.
Our discussion should then be viewed as merely suggestive of a possible holographic correspondence
between the aforementioned theories, and in order to further strengthen the case more 
observables would need to be matched. 

However, if the conditions (23) are not met  
then no viable gravity duals of the aforementioned physically revelant NFL 
systems may be found amongst the generalized family of metrics (5).  
By this argument, one concludes that none of such systems 
can be naturally mated with the classical AdS-RN metric considered in Refs.\cite{AdS2}.     

{\it Coulomb interacting Dirac fermions}

The recent upsurge of interest in graphene and topological insulators - as well the earlier advent of $1D$ Coulomb metals (e.g., carbon nanotutes), gapless $2D$ high-$T_c$ superconductors, 
and quasiparticle properties of $3D$ superfluid $He^3$ - brought out 
the problem of (pseudo)relativistic Dirac fermions with isolated Fermi points
and potentially long-ranged (due to a lack of screening), albeit nearly instantaneous, interactions.
 
In the $1D$ case, the asymptotic dual geometry is $AdS_3$ and the corresponding (conformally invariant) boundary theory is that of chiral $1D$ fermions.
Its proper UV completion is naturally achieved with 
the use of Eq.(3) for $\mu=0$, known as the 'non-rotating BTZ black hole' \cite{BTZ},
and the (exact) finite-temperature chiral fermion propagators read 
\be
G_{\pm}(\tau,x)=({\pi T\over \sinh \pi x_+T})^{2\Delta_+}
({\pi T\over \sinh(x_-T)})^{2\Delta_-}
\ee
where $x_{\pm}=x\pm\tau$.
In the $T\to 0$ limit Eq.(26) amounts to $G_{\pm}(\tau,x)=1/x_+^{2\Delta_+}x_-^{2\Delta_-}$. 

According to the holographic principle \cite{AdS}, 
the boundary theory then must be strongly coupled, 
as manifested by the UV (left/right) fermion dimensions, 
$\Delta_{\pm}=mL/2+1/2\pm 1/4$, 
which are necessarily large, 
$\Delta_{+}+\Delta_{-}>1$ \cite{BTZ}.

Notably, such dimensions can not be obtained from any $1D$ theory with short-ranged  repulsive couplings where the corresponding Luttinger parameter would be  
limited to the interval $1/2\leq K\leq 1$, thereby resulting in
$1/2\leq \Delta_{+}+\Delta_{-}={1\over 4}(K+1/K)\leq 5/8$.

In fact, the still lower $K$ values, $0<K<1/2$, can only be attained in the presence 
of long-ranged interactions, such as Coulomb, which endows the Luttinger parameter 
and the fermion dispersion with a slow momentum dependence: 
\be
K(q)={1\over {\sqrt {1+\sigma|\ln q|}}},~~~\epsilon_q=q{\sqrt {1+\sigma|\ln q|}}
\ee
where $\sigma=2e^2/\pi$.

In the general case of the $d$-dimensional Dirac fermions with 
the $3D$ Coulomb interaction, $U_q\sim q^{1-d}$,  
the counterpart of Eq.(20) reads (after the frequency integration) 
\be
S(\tau,x)=\int {d^dqU_q\over (2\pi)^d\epsilon_q}
(1-\cos(\epsilon_q\tau-qx))
\ee
In the $1D$ case, one then obtains the leading behavior 
\be
S(\tau, x)\sim\sigma^{1/2}\ln^{3/2}|x-\tau|
\ee
which gives rise to a faster-than-algebraic decay of the propagator
$G(\tau, x)\sim e^{-S}$, thus implying that the system undergoes the $1D$ analog of the Mott transition \cite{schulz}. 

Making use of Eq.(15), one observes that capturing the behavior (28) in the holographic framework would require a logarithmic deformation of the asymptotic $AdS_3$ geometry   
\be
g(z)=({L/z})^2\ln z/z_0,~~~f(z)=h(z)=({L/z})^{2}
\ee
Although this ansatz may not be unique, it shows that 
a prospective gravity dual of the Coulomb-interacting Dirac fermions 
is likely to lie outside the family of the AdS-RN metrics 
utilized in Refs.\cite{AdS2}. On the other hand, there exist 
solutions showing logarithmic behavior at intermediate values of $z$ 
for some generalized dilaton potentials, as in Eq.(4), tuned to their 
degeneracy points \cite{dilaton}. Thus, one would 
have to tap into those resources if viable candidates to the role of the 
gravity dual of the $1D$ Coulomb metal were to be found. 

{\it Summary}
 
In conclusion, we demonstrate that in order to ascertain the status of the holographic approach in the context of its applications to the conjectured NFLs one needs to venture out of the comfort zone of the customary gravitational backgrounds in search of new (self-consistent) solutions to the coupled equations for the metric, dilaton, gauge, and matter fields along the lines of Refs.\cite{lifshitz,dilaton,electronstar,hyperscaling}.

To that end, we discuss the specific examples of non-relativistic fermions coupled to gauge-like fields and Dirac fermions with the Coulomb interactions, in both cases finding the prospective metrics to possibly belong to the solutions
(alongside the hyperscaling-violating ones) of some generalized dilaton gravity models. 

Based on these observations, we suggest that within such a broad class of metrics one would have better chances of 'reverse engineering' the gravity duals of the already documented NFLs, 
thus putting the entire holographic machinery up to a decisive test. 
Then, after having affirmed that this approach might work in the already known cases, 
one can continue expanding the list of novel NFL states with a greater confidence.

%\begin{references}

\end{document}